\documentclass[aps,preprint,groupedaddress,superscriptaddress]{revtex4}

\usepackage{amssymb,amsfonts,amsmath}
\usepackage{adjustbox}
\usepackage{hyperref}
\usepackage{color}
\usepackage{xr}
\usepackage{braket}
\usepackage{xcolor}
\usepackage{float}
\usepackage{setspace}
\usepackage{ulem}
\usepackage{subfigure}
\usepackage{soul}

\begin{document}

\title{Optimizing Density Functional Theory for Strain-Dependent Magnetic Properties of Monolayer MnBi\(_2\)Te\(_4\) with Diffusion Monte Carlo}

\author{Jeonghwan Ahn}
\email{kindazet@gmail.com}
\affiliation{Materials Science and Technology Division, Oak Ridge National Laboratory, Oak Ridge, Tennessee 37831, USA}

\author{Swarnava Ghosh}
\affiliation{National Center for Computational Science, Oak Ridge National Laboratory, Oak Ridge, Tennessee 37831, USA}

\author{Seoung-Hun Kang}
\affiliation{Materials Science and Technology Division, Oak Ridge National Laboratory, Oak Ridge, Tennessee 37831, USA}

\author{Dameul Jeong}
\affiliation{Department of Physics, Kyung Hee University, Seoul, 02447, South Korea}
\author{Markus Eisenbach}
\affiliation{National Center for Computational Science, Oak Ridge National Laboratory, Oak Ridge, Tennessee 37831, USA}

\author{Young-Kyun Kwon}
\affiliation{Department of Physics, Kyung Hee University, Seoul, 02447, South Korea}
\affiliation{Department of Information Display and Research Institute for Basic Sciences, Kyung Hee University, Seoul, 02447, South Korea}

\author{Fernando A. Reboredo}
\affiliation{Materials Science and Technology Division, Oak Ridge National Laboratory, Oak Ridge, Tennessee 37831, USA}

\author{Jaron T. Krogel}
\affiliation{Materials Science and Technology Division, Oak Ridge National Laboratory, Oak Ridge, Tennessee 37831, USA}

\author{Mina Yoon}
\email{myoon@ornl.gov}
\affiliation{Materials Science and Technology Division, Oak Ridge National Laboratory, Oak Ridge, Tennessee 37831, USA}

\begin{abstract}
Monolayer MnBi$_{2}$Te$_{4}$ (MBT) is an intrinsically magnetic topological insulator whose magnetic response is strongly affected by strain and electron correlation. 
In density functional theory with an on-site Hubbard correction (DFT+$U$), however, predictions vary substantially with the choice of Hubbard $U$, making it difficult to establish a reliable strain-dependent picture of magnetism in this system.
Here we use diffusion Monte Carlo (DMC) to benchmark DFT+$U$ for monolayer MBT and to determine an effective $U$ as a function of strain.
We find that the predicted magnetic phase diagram depends strongly on $U$, indicating that a single fixed value is not sufficient across the strain range considered. 
DMC nodal optimization further shows that the optimal $U$ increases with strain magnitude and is well captured by a simple quadratic form. 
When this DMC-informed strain-dependent $U$ is used in PBE+$U$, the calculated Mn local moments are brought into close agreement with DMC and are improved relative to commonly used fixed-$U$ choices.
These results show that, for monolayer MBT, correlation strength itself should be treated as strain dependent, and they provide a practical many-body-guided strategy for improving strain-dependent DFT+$U$ descriptions of magnetic van der Waals materials.

\end{abstract}

\onecolumngrid
Notice: This manuscript has been coauthored by UT-Battelle, LLC, under Contract No. DE-AC0500OR22725 with the U.S. Department of Energy. The United States Government retains and the publisher, by accepting the article for publication, acknowledges that the United States Government retains a non-exclusive, paid-up, irrevocable, world-wide license to publish or reproduce the published form of this manuscript, or allow others to do so, for the United States Government purposes. The Department of Energy will provide public access to these results of federally sponsored research in accordance with the DOE Public Access Plan (http://energy.gov/downloads/doe-public-access-plan).

\maketitle

\section{Introduction}
MnBi$_{2}$Te$_{4}$ (MBT) is a prototypical magnetic van der Waals (vdW) topological material in which magnetic order and nontrivial band topology are intimately coupled~\cite{ zhang2009topological, konig2007quantum, chen2009experimental, hasan2010colloquium, kane2005z, fu2007topological, qi2008topological, xiao2018realization, chen2010massive,otrokov2019prediction,chang2013experimental,li2019intrinsic, he2020mnbi2te4, zhang2019topological, deng2020quantum}. 
The MBT family consists of septuple layers stacked in the sequence Te-Bi-Te-Mn-Te-Bi-Te, and its magnetic behavior depends strongly on thickness, stacking environment, and external perturbations such as temperature and strain~\cite{otrokov2019prediction,li2019intrinsic, he2020mnbi2te4,klimovskikh2020tunable, yan2022perspective, guo2021pressure, ovchinnikov2021intertwined,xue2020control, gao2021mbt, lei2021metamagnetism}.
In the two-dimensional limit, monolayer MBT is especially attractive because it isolates intralayer magnetism from the additional complexity of interlayer magnetic coupling present in multilayers and bulk crystals. 
This makes the monolayer a useful platform for examining how structural perturbations influence correlation-sensitive magnetism.

Despite extensive theoretical and experimental work on MBT, several aspects of its electronic and magnetic behavior remain under debate~\cite{yan2022perspective, ahn2023mbt}. 
In particular, discrepancies between experimental observations and first-principles predictions have complicated the interpretation of its magnetic and topological properties~\cite{yan2022perspective, ahn2023mbt,li2019dirac, hao2019gapless, chen2019topological, swatek2020gapless, nevola2020coexistence}. 
While some of these discrepancies may be related to structural imperfections such as stacking faults in multilayer samples~\cite{ahn2023mbt}, they also point to a broader theoretical issue. 
The predicted magnetic energetics of MBT are highly sensitive to how electron correlation on the Mn 3d shell is treated within density functional theory (DFT).
Within DFT plus an on-site Hubbard $U$ correction (DFT+U), Hubbard $U$ values ranging from about 3 to 5.34 eV have been used in previous studies of MBT~\cite{otrokov2019prediction, klimovskikh2020tunable, xue2020control, hao2019gapless, zhu2021tunable, lai2021defect, du2021tuning}, often leading to substantially different conclusions regarding magnetic stability. 
Since the magnetic order is closely tied to the topological response of this material family, establishing a better-founded description of correlation effects is essential~\cite{luo2023mbt}.

Strain provides a particularly useful way to examine this problem. 
Previous studies have shown that strain can modify magnetic couplings, phase boundaries, and magnetic moments in MBT and related systems~\cite{klimovskikh2020tunable,yan2022perspective, guo2021pressure, ovchinnikov2021intertwined,xue2020control, gao2021mbt, lei2021metamagnetism}.
From both fundamental and applied perspectives, strain is therefore a natural control parameter for probing the interplay between lattice distortion and magnetism in a magnetic vdW material. 
At the same time, strain is precisely the kind of perturbation for which a fixed empirical $U$ may become unreliable, because the degree of orbital overlap and localization can change with structure. 
A central question is therefore whether a single $U$ value can describe monolayer MBT across different strain conditions, or whether the effective correlation strength itself should be treated as strain dependent.

To address this question, a benchmark beyond standard DFT is needed. 
Diffusion Monte Carlo (DMC) provides a many-body reference that is well suited for assessing correlation-sensitive properties~\cite{foulkes2001}. 
In fixed-node DMC, the nodal surface depends on the underlying trial wave function, which in the present context can be generated from DFT+$U$ orbitals.
This makes it possible to use DMC as a variational benchmark for identifying an optimal $U$ for a given structure and for tracking how the effective optimal $U$ evolves with strain. 
Rather than treating $U$ as a fixed empirical input, this approach ties it directly to a many-body criterion.

In this work, we investigate the strain-dependent magnetic properties of monolayer MBT using DFT+$U$ and DMC. 
We first show that the predicted magnetic phase diagram depends strongly on the choice of Hubbard $U$, indicating that correlation treatment is a central source of theoretical uncertainty. 
We then use DMC nodal optimization to determine the optimal $U$ as a function of strain and construct a simple strain-dependent model for this quantity.
Finally, we use this DMC-informed $U$ within the DFT+$U$ framework to re-evaluate the magnetic response of monolayer MBT under strain. 
Our results identify monolayer MBT as a useful two-dimensional testbed for benchmarking correlation-sensitive magnetism and provide a practical route for improving strain-dependent DFT+$U$ predictions, particularly for magnetic moments, in magnetic vdW materials.

\section{Computational Approaches}
\subsection{Density Functional Theory}
We utilized the Vienna {\it{ ab initio}} simulation package (VASP) based on density functional theory to carry out a first-principles study and compute the magnetic properties of MBT~\cite{{hohenberg1964inhomogeneous,kohn1965self},kresse1993ab,kresse1996efficient}. To ensure convergence, we adopted a plane wave cutoff of 400~eV with projector augmented wave method (PAW) pseudopotentials~\cite{blochl1994projector} and employed an $18\times18\times1$ $\Gamma$-centered $k$-points sampling with a convergence criterion of 0.05~meV in magnetic anisotropy energy for pristine MBT. 

As exchange-correlation functional, generalized gradient approximation (GGA) of Perdew-Burke-Ernzerhof (PBE)~\cite{GGA}, and Ceperley-Alder form~\cite{ceperley1980ground} within the local density approximation (LDA) was used. We set the Hellmann-Feynman force criterion to 0.01~eV/$\AA$ for structural and atomic relaxation and included spin-orbit coupling (SOC) interaction in all calculations. Mn has partially filled $d$-orbitals, and their electronic behavior can be intricate, involving strong electron-electron interactions. Hence, in such cases, the introduction of Hubbard $U$ in the framework of DFT is necessary to account for the on-site Coulomb repulsion between electrons in the transition metal $d$-orbitals. We studied the magnetic properties of MBT using the $U$ parameter values between 0~eV and 4.4~eV on Mn 3$d$ orbitals, and  the DMC optimized values for the given strain were used to evaluate the accurate magnetic anisotropy energy.

For selected cases, we compared the electronic ground states using the all-electron method as implemented in Elk~\cite{elk} and the pseudopotential method as implemented in VASP~\cite{kresse1993ab, kresse1996efficient}. 
In Elk, the DFT+$U$ method is implemented using either the around mean-field (AMF) \cite{amf} or fully localized limit (FLL) \cite{fll} schemes. This comparison allows us to evaluate the impact of different DFT+$U$ implementations on the magnetic properties of MBT, providing a more comprehensive understanding of the impact of different approximations of the theory under different computational approaches. Readers are referred to the section I of Supplementary Information for details and discussion of the all-electron calculations.

\subsection{Diffusion Monte Carlo}
DMC is a real-space method to directly solve the many-body Schr\"{o}dinger equation based on the stochastic process. 
DMC calculations were implemented with QMCPACK~\cite{kim18,kent2020} code under the fixed-node approximation~\cite{anderson1975,anderson1976} that makes the projected state remain antisymmetric as a fermionic system.
Because the Hubbard $U$ can serve as a variational parameter in optimizing the nodal surface used in the fixed-node DMC calculations, one can find an optimal value of Hubbard $U$ to yield the lowest DMC energy for the given structure.
This allows us to model values of Hubbard $U$ depending on the strain of the system, as discussed later. 
We employed the Slater-Jastrow type function for a many-body trial wave function. The Slater determinants were constructed with DFT Kohn-Sham~\cite{kohn1965self} orbitals based on the PBE+$U$ functional
where the corresponding DFT calculations were performed with QUANTUM ESPRESSO~\cite{gianozzi2009} (QE) package. 

We considered one-, two- and three-body Jastrow factors to describe electron-ion, electron-electron, and electron-electron-ion correlation effects, respectively. 
The norm-conserving pseudopotentials developed within ccECP approach~\cite{Bennett2017,Bennett2018} were used to represent the atomic cores as in our previous DMC-informed DFT studies for the MBT system~\cite{ahn2025diffusion}.
The Jastrow optimizations were done with variational Monte Carlo calculations based on the linear method of Umrigar {\it et al.}~\cite{umrigar2007} and subsequent DMC calculations were done using a time step of $\tau = 0.005$ Ha$^{-1}$. To evaluate non-local pseudopotentials, imaginary-time projections were treated with size-consistent T-moves~\cite{Casula2010}. 
The fixed-node DMC calculations for the nodal surface optimization were performed with 7-atom supercell structure as in our previous DMC calculations on bulk MBT~\cite{ahn2025diffusion}, showing that there is little difference between the DMC results of 7-atom and 28-atom supercells.

All DMC calculations employed orbitals generated from Quantum ESPRESSO using the FLL for the DFT+$U$ scheme. 
Our VASP calculations also used the FLL double-counting correction with atom-centered projectors within the PAW framework. 
Despite minor code-specific differences in pseudopotential formats and projector normalization, both implementations follow the same formalism, ensuring comparable charge localization properties.
Therefore, the DMC-optimized $U$ values obtained with QE orbitals are physically transferable to the VASP framework, justifying the use of a common “DMC-informed” U parameter across codes.
Importantly, we do not use the raw magnetic moments directly from the DFT outputs. Instead, the local moments are consistently evaluated using the procedure described below, so that DFT and DMC values share the same definition and are directly comparable, following the correction protocol reported in Ref.~\cite{ahn2025diffusion}.

To assess the Mn local moments in MBT, we evaluate the spin polarization around each Mn site for both FM and A-AFM configurations using DMC and DFT. The spin density, $\rho^s(\mathbf{r})$, is obtained from the difference between the spin-resolved electron densities, $\rho^{\uparrow}(\mathbf{r})$ and $\rho^{\downarrow}(\mathbf{r})$, as $
\rho^s(\mathbf{r}) = \rho^{\uparrow}(\mathbf{r})-\rho^{\downarrow}(\mathbf{r}).$
The site-resolved magnetic moment $M$ is defined by integrating $\rho^s(\mathbf{r})$ within a sphere of radius $R$ centered at atomic site $I$,
\begin{align}
    \begin{split} 
        M_I(R) &= \int_0^Rdr \int_0^{\pi}d\theta\int_0^{2\pi}d\phi \,\rho^s(\mathbf{r}-\mathbf{r}_I)
    \end{split} 
\end{align}
and the radius $R$ is increased until $M_I(R)$ converges to a plateau value as seen in Ref.~\cite{ahn2025diffusion}.

Because electron density operators do not commute with the Hamiltonian, the mixed distribution sampled in DMC, $\Psi_{T}(R)\Phi_{\text{FN}}(R)$, yields a biased estimate proportional to the difference $\Psi_{T}-\Phi_{\text{FN}}$, where $\Psi_{T}$ is the trial function and $\Phi_{\text{FN}}$ the fixed-node ground state. 
To reduce this bias we use the extrapolated estimator~\cite{foulkes2001}, $\rho_{\text{ext}} = 2\rho_{\text{DMC}}-\rho_{\text{VMC}}$, where $\rho_{\text{VMC}}$ is the variational Monte Carlo density.

\section{Results}
\subsection{Hubbard-$U$ dependent magnetic states}
%
\begin{figure}[t]
\includegraphics[width=0.7\linewidth, angle=0] {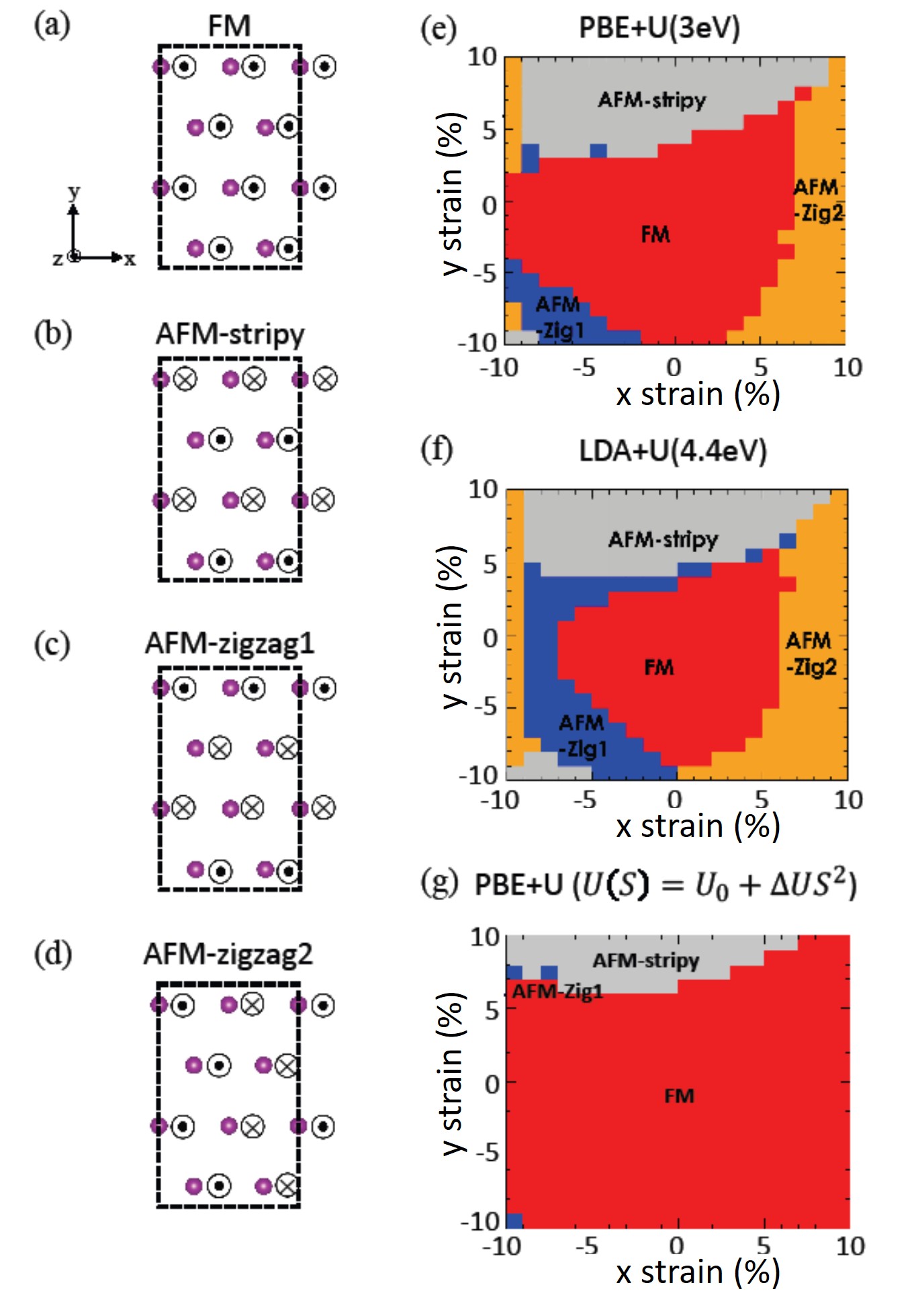}
\caption{Magnetic configurations of the monolayer MBT considered in this study: FM (a), AFM-stripy (b), AFM-zigzag 1 (c), and (d) AFM-zigzag 2. The stable magnetic phases among these four magnetic states depending on the in-plane strain, x and y, up to 10\% of the tensile and compressive strains. The magnetic phase diagram changes drastically depending on the choice of exchange-correlation functional and Hubbard $U$ values: (e) PBE+$U$ = 3~eV, (f) LDA+$U$ = 4~eV, and (g) PBE+$U_{DMC}$, DMC optimized $U$ depending on strain.}
\label{mag_phase_diagram}
\end{figure} 
%
Using the DFT+U approach we study the influence of the Hubbard $U$ value  on the magnetic ground state in monolayer  MBT. Figure~\ref{mag_phase_diagram} illustrates the magnetic configurations of the monolayer MBT considered in this study - they are ferromagnetic (FM) and three different antiferromagnetic (AFM) configurations forming a stripe pattern, and two different types of zigzag ordering of spin up and down components, either along $x$ or $y$ directions. The stable magnetic phases among these four magnetic states depend on the in-plane strain (on the $xy$ plane), up to 10\% of the tensile and compressive strains. 
Most first-principles studies of MBT have adopted Hubbard $U$ values in the range of approximately 3 to 5 eV~\cite{zhu2021tunable, lai2021defect, du2021tuning, xue2020control, hao2019gapless, otrokov2019prediction, klimovskikh2020tunable}, often leading to substantially different predictions for magnetic stability.

Our calculations show that the magnetic phase diagram is highly sensitive to both the choice of exchange-correlation functional, namely local density approximation (LDA) versus generalized gradient approximation (GGA), and the value of the Hubbard $U$.
In the absence of strain, DMC yields an optimal $U$ of 4.0(2) eV for monolayer MBT. 
For context, previous bulk benchmarks gave 4.4 eV for LDA+$U$ and 3.5 eV for PBE+$U$. 
If these values are then held fixed across strain, the predicted magnetic phase diagram changes significantly depending on the chosen functional and $U$.
For example, PBE+$U$ =3 eV calculations predict that FM states are favorable over a wide range of $x$-strain, while LDA+$U$ =4.4 eV calculations show a much larger range of AFM zigzag1 phases and a smaller range of FM phases in the strain range (see Fig.~\ref{mag_phase_diagram}). 
The strain-dependent DMC-optimized values, $U_{\text{DMC}}$, discussed below, yield a qualitatively different phase diagram (see Fig.~\ref{mag_phase_diagram}(g)), further illustrating the strong sensitivity of the strain-dependent magnetic landscape to the treatment of $U$.

%
\begin{figure}
\centering
\subfigure[GGA+U]{\includegraphics[width=0.4\linewidth, angle=0] {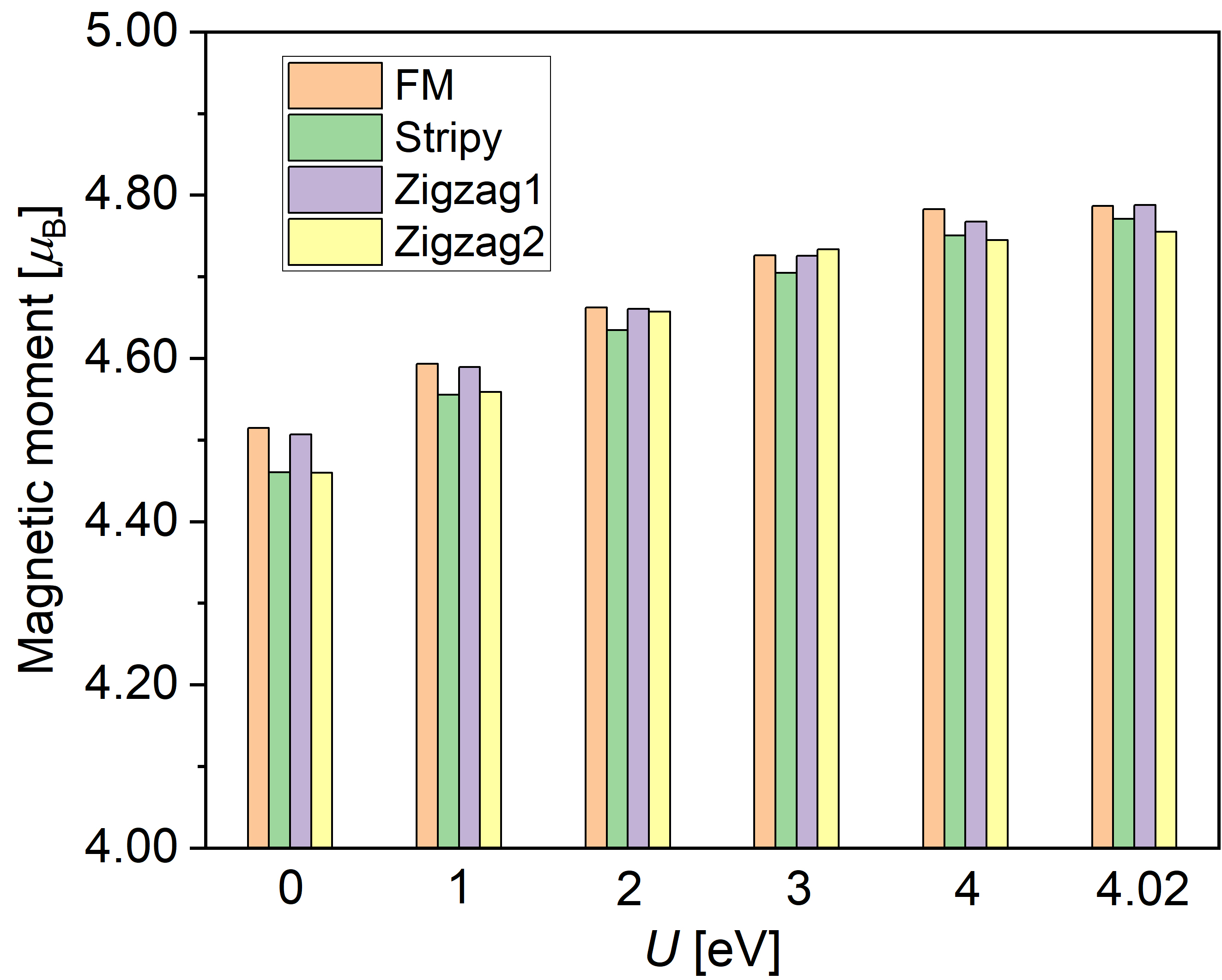}}
\subfigure[LDA+U]{\includegraphics[width=0.4\linewidth, angle=0] {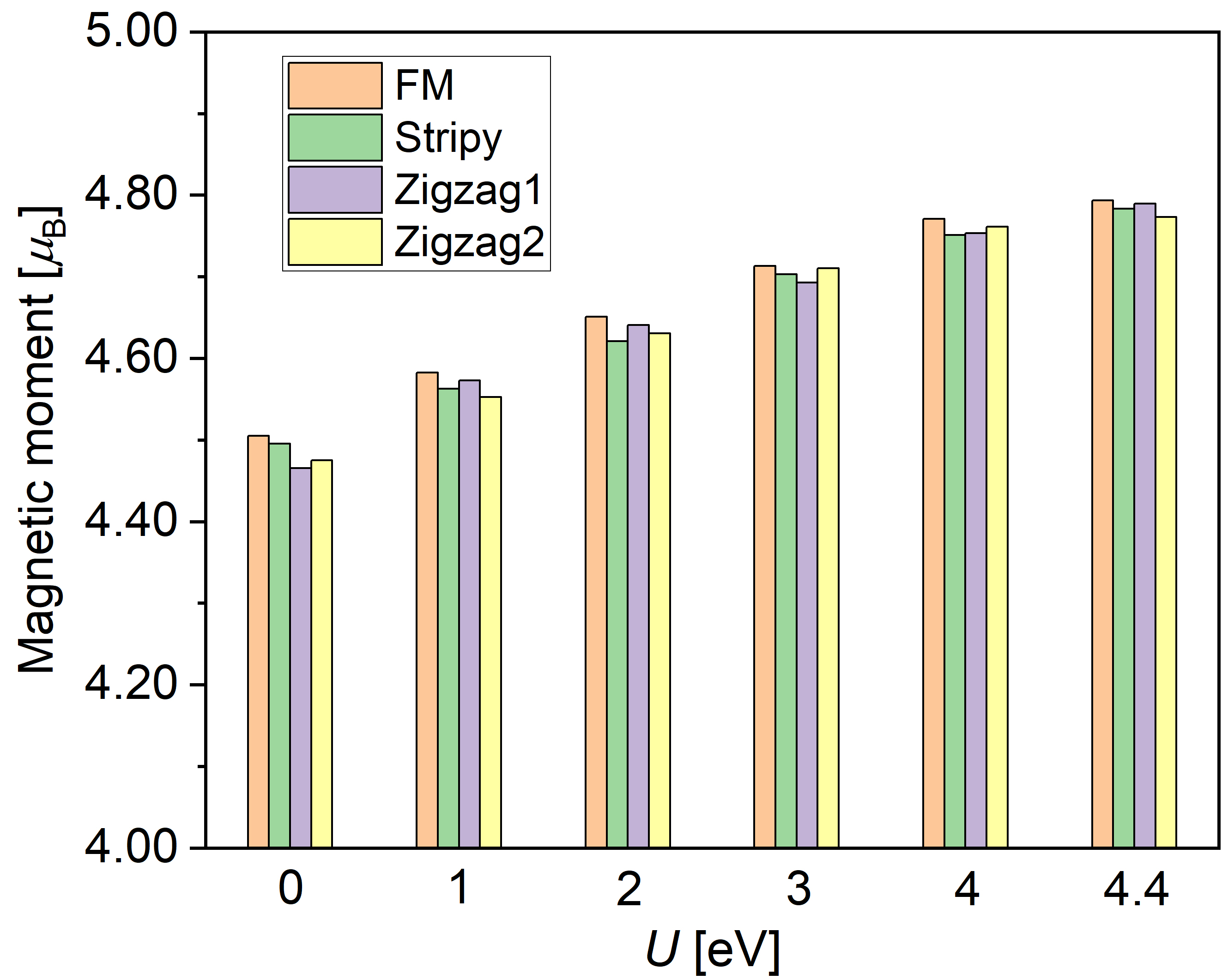}}
\caption{Magnetic moments of the Mn atoms in the pristine (unstrained) monolayer MBT based on the calculations with a) GGA+$U$ and b) LDA+$U$.} 
\label{moment_U}
\end{figure} 
%

In contrast to the phase diagram, the Mn local moment depends only moderately on $U$ and becomes nearly saturated above about 3 eV.
Figure \ref{moment_U} shows the magnetic moments at Mn atoms for different values of $U$ for a) GGA+$U$ and b) LDA+$U$ cases. 
The value of the magnetic moment increases from 4.45 $\mu_B$ to 4.78 $\mu_B$ as $U$ is increased from 0 to 4.02 eV in the GGA+$U$ case and from 4.47 $\mu_B$ to 4.79 $\mu_B$ as $U$ is increased from 0 eV to 4.4 eV in the LDA+$U$ case. 
Spin-orbit coupling and magnetic configuration (FM, stripy, zigzag 1, and zigzag 2) have a negligible effect on the magnetic moment.

%
\begin{figure}
\begin{center}
\subfigure[GGA+U]{\includegraphics[width=0.4\linewidth, angle=0] {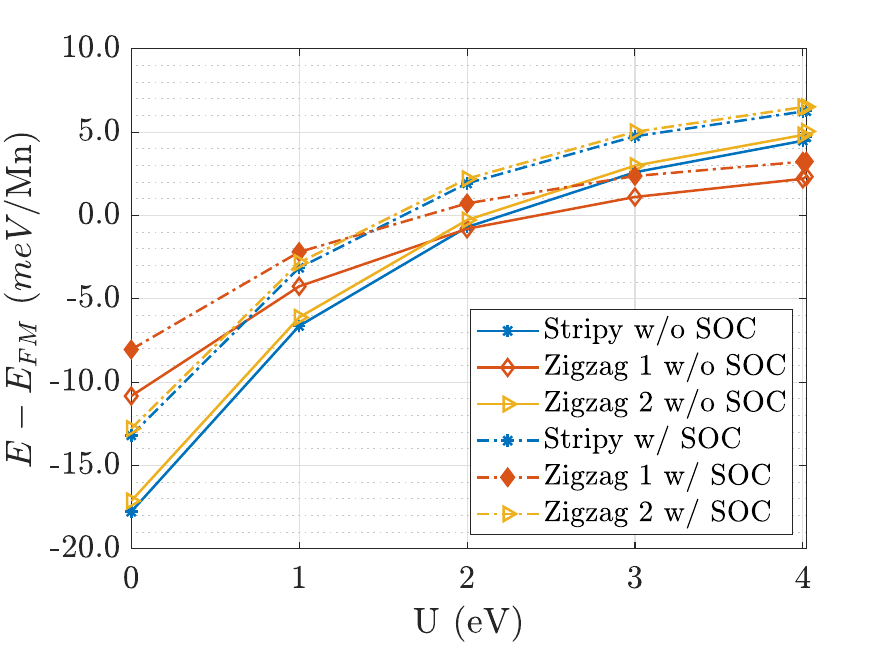}}
\subfigure[LDA+U]{\includegraphics[width=0.4\linewidth, angle=0] {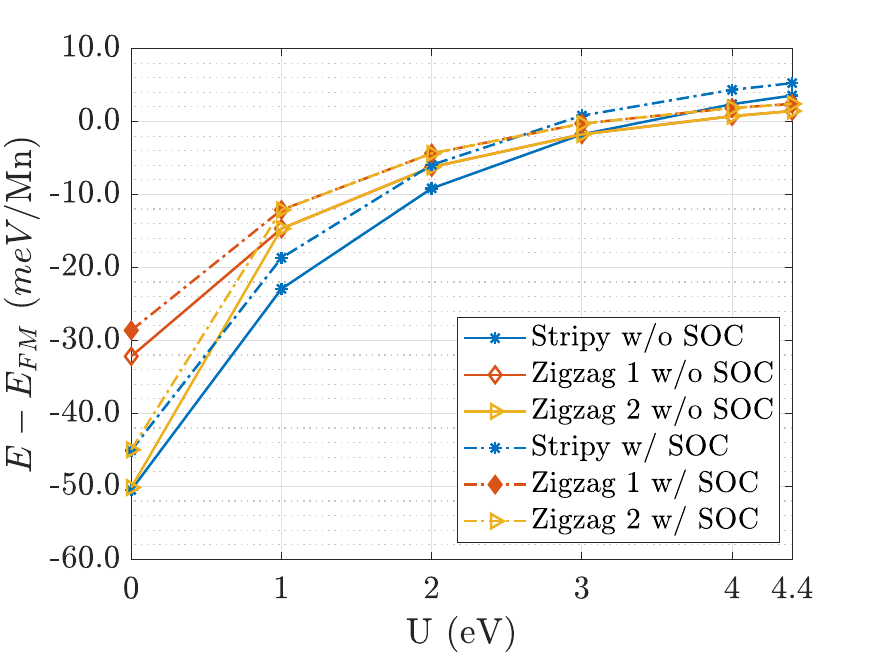}}
\caption{ Energy difference versus different choices of the $U$ parameter for a) GGA+$U$ and b) LDA+$U$ exchange correlation functionals. The energy difference is defined as the difference between the different magnetic states and the ferromagnetic state. } 
\label{energy_U}
\end{center}
\end{figure} 
%
Next, we study the influence of the Hubbard $U$ on the energy difference between different magnetic states (see Fig. \ref{mag_phase_diagram}). The supercell has been optimized for each value of $U$. Fig. \ref{energy_U} shows the energy difference as a function of $U$ for the GGA+$U$ and LDA+$U$ calculations. The energy difference is defined as the difference between the magnetic state and the ferromagnetic state ($E-E_{FM}$). From Fig. \ref{energy_U}a we observe that for $U$ = 0 eV, AFM stripy is the favored magnetic ground state for both GGA and LDA exchange correlations. When the value of $U$ is further increased to 4.02 eV for GGA+$U$ and 4.4 eV for LDA+$U$, the favored magnetic ground state changes to an FM state. The transformation from AFM stripy to FM occurs between $U$ = 2-3 eV for GGA+$U$ and 3-4 eV for LDA+$U$. Note also that the relative ordering of the magnetic states is different for the GGA+$U$ and LDA+$U$ cases. For GGA+$U$ = 4.02 eV, $E_{FM} < E_{zigzag 1} < E_{stripy} < E_{zigzag 2} $, and for LDA+$U$ = 4.4 eV, $E_{FM} < E_{zigzag 1} = E_{zigzag 2} < E_{stripy} $. 
For GGA+$U$, both QE and VASP predict the same ordering of the magnetic phases. However, QE yields larger relative energy differences compared to VASP, which can shift the phase boundaries and lead to a stronger stabilization of the FM phase (See Table I in section II of Supplementary Information).

We conclude that a too small $U$ value leads to an underestimation of the magnetic moment, which is known from neutron scattering experiments to be about 4.7(1) $\mu_B$~\cite{Ding_PhysRevB_2020} in bulk MBT. 
Thus, too small a $U$ underestimates the magnetic moment and erroneously favors non-ferromagnetic ordered states at zero strain.
Most of the differences between LDA+$U$ and PBE+$U$ can be approximately understood as reflecting an effective shift of about 1-1.5 eV in $U$.
These results highlight the importance of determining the Hubbard $U$ parameter accurately when describing the magnetic properties and phase behavior of MBT under varying conditions. 
While the DMC energy minimization approach provides an optimized $U$ parameter that improves the consistency of magnetic moment predictions, we recognize that its relationship to the energetics of magnetic states is more tenuous, since the DMC energy depends primarily on the quality of the nodal surfaces. This approach should therefore be seen as a pragmatic method for improving DFT+$U$ predictions, particularly for magnetic moments, rather than as a comprehensive theoretical framework.

\subsection{Determining optimized-$U$ depending on strain via DMC}
\label{sec:dmc_u_model}

%
\begin{figure}[h]
\centering
\includegraphics[width=0.4\linewidth] {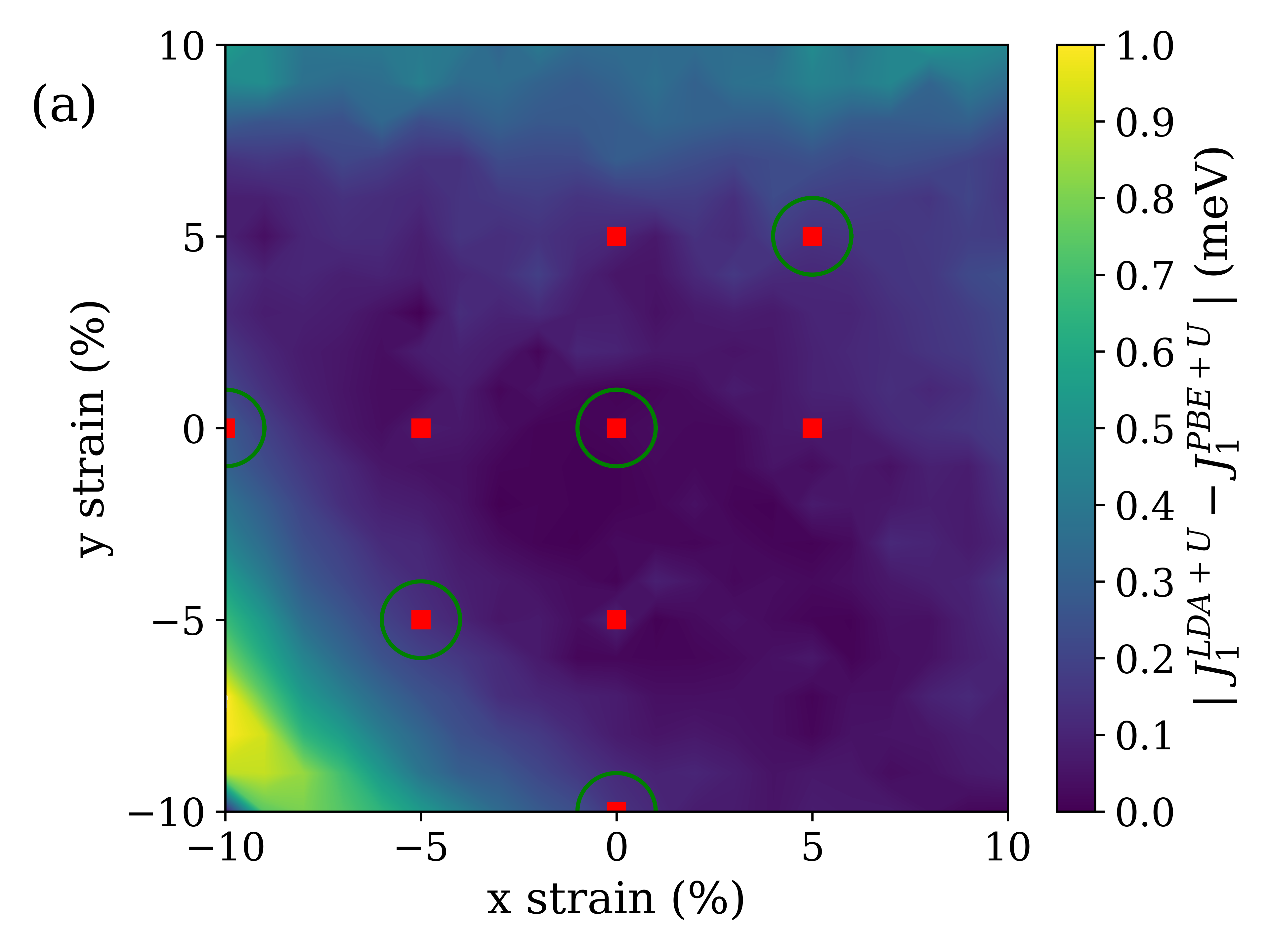}
\includegraphics[width=0.4\linewidth] {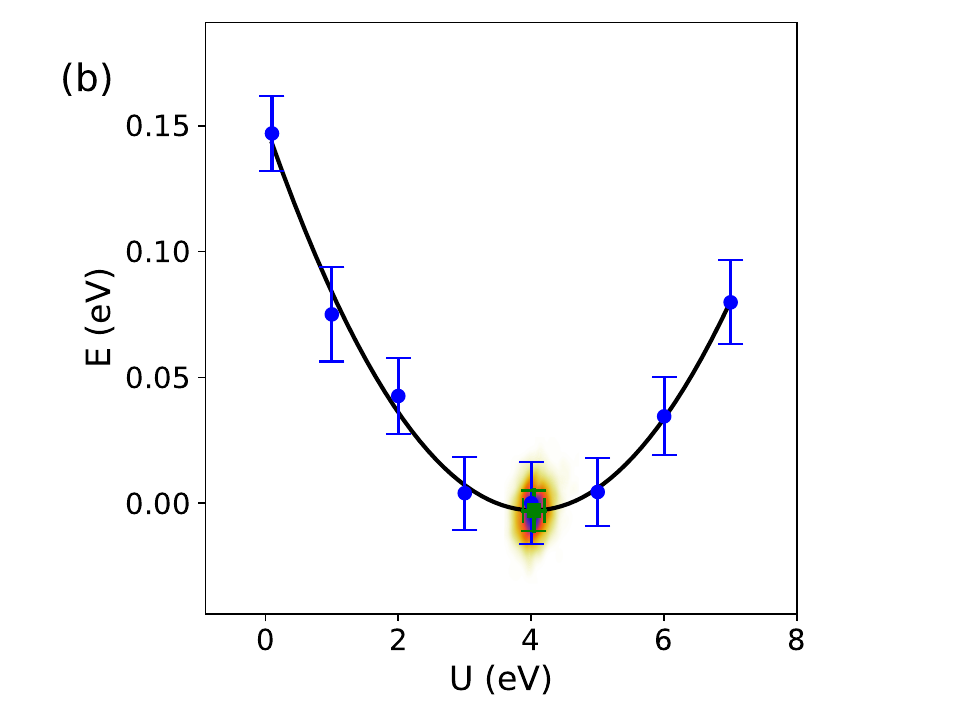}
\caption{Sensitivity analysis to guide optimal determination of Hubbard U via DMC.  In (a) the absolute difference in $J_1$ exchange parameter between LDA+$U$ = 4.4 eV and PBE+$U$ = 3 eV is used to indicate the spin model sensitivity to U.  Strain regions showing high sensitivity (large absolute $J_1$ difference) are used to select $x$ and $y$ strain points (circled in green) for subsequent determination of optimal $U$ via DMC.  In (b) the DMC optimized $U$ (for PBE+$U$) is determined by the variational principle at zero strain.  The distribution of minima resulting from resampled quadratic fits are shown in color. } 
\label{fig:dmc_design}
\end{figure} 
%

Because strain is expected to alter the magnetic properties of MBT, it is important to understand the relationship between strain and Hubbard $U$.  
For this purpose, we use highly accurate diffusion Monte Carlo (DMC) methods to determine 
the optimal $U$ for PBE+$U$ and how it varies with strain in the monolayer MBT.
Since DMC is computationally demanding, we first performed a sensitivity analysis based 
on the dominant exchange contribution ($J_1$) that enters into spin models for the magnetic 
behavior of this material.  
As a starting point, we evaluated $J_1$ as a function of strain using two representative functionals motivated by previous bulk studies: PBE+$U$ = 3 eV and LDA+$U$ = 4.4 eV. 
A previous DFT study employed PBE+$U$ = 3 eV for bulk MBT, whereas a prior DMC benchmark identified LDA+$U$ with $U$ = 4.4 eV as optimal for the same system. We therefore computed $J_1$ as a function of strain using both choices.
In Fig. \ref{fig:dmc_design}a we show the absolute difference in $J_1$ calculated with these two functionals as a function of strain for the monolayer MBT.  Large absolute differences indicate strain regions that are sensitive to the underlying assumptions of the DFT functionals, and thus serve to identify where better information about the functional is most needed.  

Based on this, we selected four strain points (highlighted by green circles in Fig. \ref{fig:dmc_design}a), in addition to the origin, to perform nodal optimization as a function of $U$ for PBE+$U$ with DMC.  These points were $(0 \%,0\%),~(5 \%,5\%),~(-5 \%,-5\%),\\
~(-10 \%,0\%),~(0 \%,-10\%)$ x,y strain. The resulting nodal optimization for zero strain is shown in Fig. \ref{fig:dmc_design}b.
Quadratic fits were performed based on resampled DMC energy data to produce a best estimate, including statistical uncertainty for the optimal $U$.  The independent estimate for $U$ at zero strain is $4.04(16)$ eV, which may differ slightly from the previously determined optimal $U$ value of 3.5(2) eV for PBE+$U$ in MBT bulk~\cite{ahn2025diffusion}.

%
\begin{figure}
\centering
\includegraphics[width=0.4\linewidth] {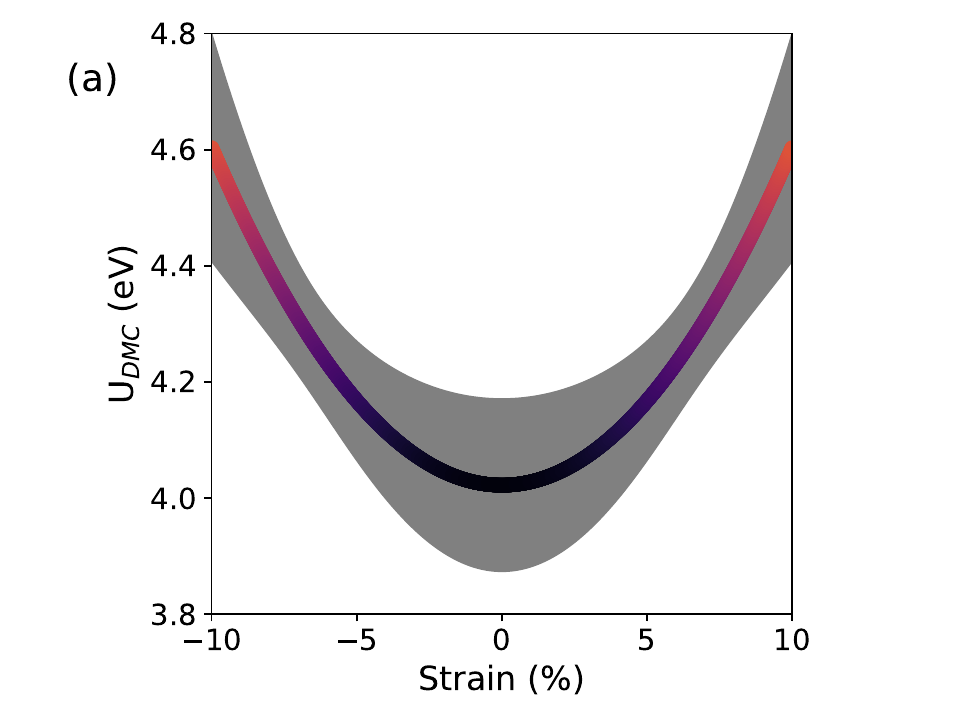}
\includegraphics[width=0.4\linewidth] {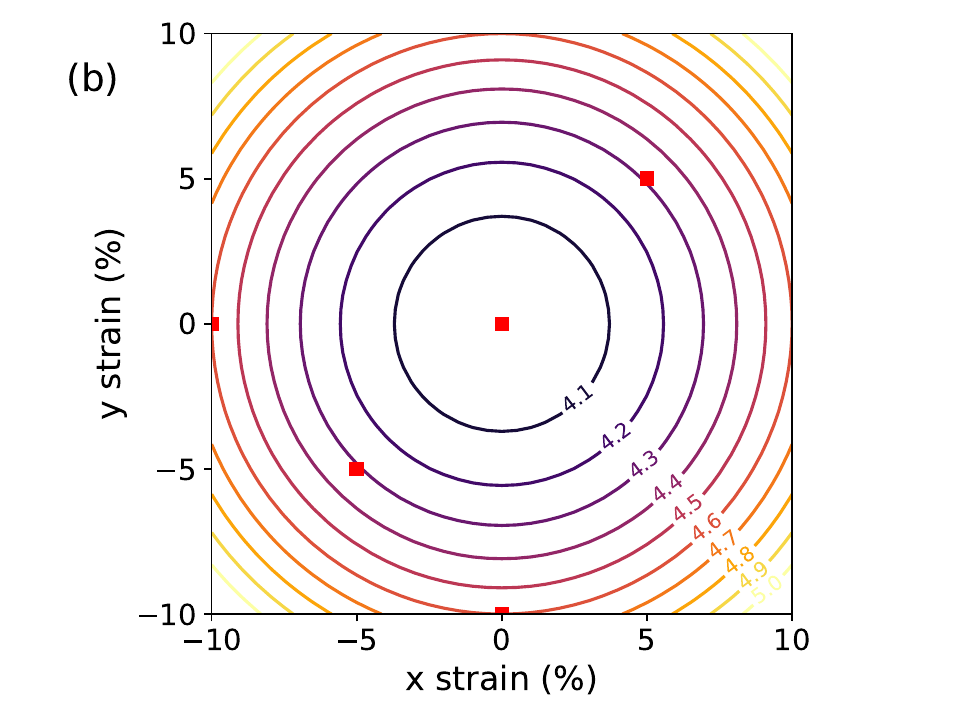}
\caption{Variation in DMC optimized U vs strain.  Panel (a) shows the near isotropic strain model for U supported by the DMC data collected at multiple strains ($U_{\mathrm{DMC}}(S)=U_0+\Delta U S^2$), with $S$ being the absolute magnitude of total strain.  Panel (b) shows the same model for U, but as contours over the x/y strain field of the monolayer MBT.} 
\label{fig:dmc_U_model}
\end{figure} 
%

The resulting optimal U values for all strain points (Fig.~\ref{fig:dmc_U_model}) were then combined to arrive at a simple model for the dependence of $U$ on strain for monolayer MBT.  For all values of strain, we found an increase in the optimal $U$.  The simplest model supported by these data with analyticity at zero strain is a simple quadratic form.  Since the difference in optimal $U$ values between points with identical total strain $S=\lvert\lvert\vec{S}\rvert\rvert$ was statistically indistinguishable from zero, an isotropic model was used:
\begin{equation}
U_\mathrm{DMC}(S) = U_{0}+\Delta{U}S^2,
\label{udmc}
\end{equation}
Based on this model, we found fitting parameters of $U_0=4.02(15) eV$ and $\Delta U=0.0058(29)$ eV.  The resulting model, with uncertainty shown as a grey background, appears in Fig. \ref{fig:dmc_U_model}a.  The same model is shown over the 2D strain field in Fig. \ref{fig:dmc_U_model}(b) with contours showing the $U$ values for each value of strain.

%
\begin{figure}
\centering
\includegraphics[width=0.7\linewidth] {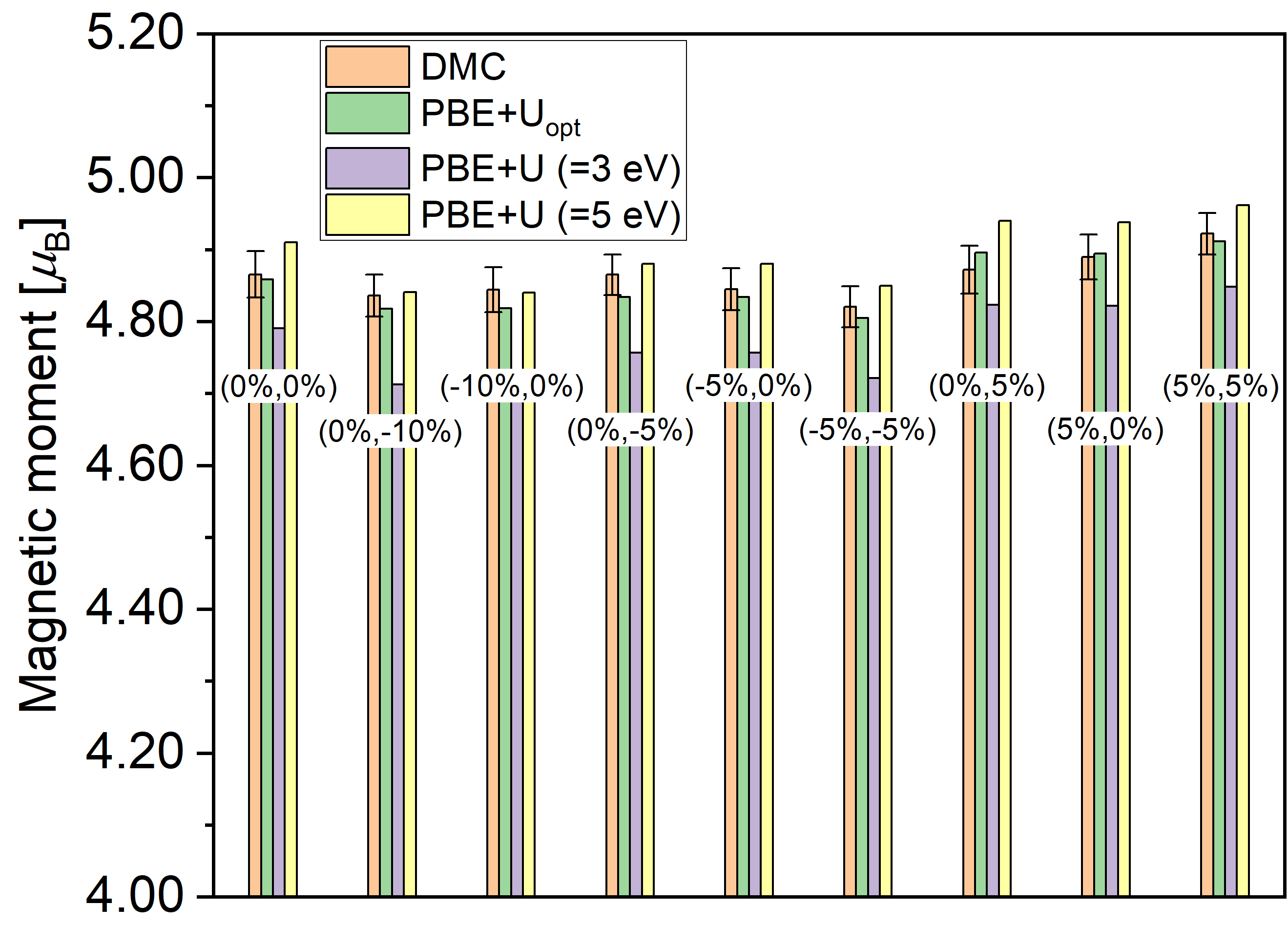}
\caption{Magnetic moments of the Mn atoms in the strained and pristine monolayer MBT computed with DMC and PBE + DMC-optimized $U$ depending on the strain. The results with PBE+$U$ =3 eV and PBE+$U$ = 5 eV are also presented for comparison.}
\label{fig:dmc_pbe_comparison}
\end{figure} 
%

Based on this model, we computed the magnetic moments of monolayer MBT under various strain conditions, as shown in Fig.~\ref{fig:dmc_pbe_comparison}. The DMC calculations for pristine monolayer MBT yield a magnetic moment of 
4.86(3) $\mu_{\text{B}}$. Tensile strain slightly increases the magnetic moment, while compressive strain slightly decreases it; however, these variations remain within the statistical error bars. Notably, the magnetic moments of strained monolayer MBT are statistically identical to those of bulk MBT and consistent with experimental values. This finding demonstrates that the magnetic moments of Mn atoms in MBT are robust against variations in interlayer stacking and strain. These results further support the use of DMC as a reliable reference for calibrating strain-dependent $U$ in monolayer MBT. Furthermore, PBE calculations combined with strain-dependent Hubbard U values optimized from DMC data reproduce the DMC magnetic moments with a root-mean-square error (RMSE) of only 0.013 $\mu_{B}$, whereas fixed-U calculations (3 and 5 eV) yield substantially larger deviations (RMSEs of 0.081 $\mu_{B}$ and 0.038 $\mu_{B}$, respectively), falling outside the DMC prediction range.
This suggests that PBE calculations informed by strain-dependent $U$ can provide more reliable predictions for the magnetic properties of monolayer MBT than fixed-$U$ calculations. 
We therefore view this approach as a pragmatic way to improve DFT+$U$ predictions, particularly for magnetic moments, rather than as a comprehensive theoretical framework.

Although the model used here is isotropic, we point out that its impact on the magnetic properties is not.  Based on the sensitivity analysis performed as shown in Fig. \ref{fig:dmc_design}, its main impact is to alter $J_1$ along the biaxial compressive and uniaxial compressive (y-axis) strain directions.
Indeed, as shown in Fig.~\ref{mag_phase_diagram}, the magnetic phase diagram derived from PBE calculations with strain-dependent $U$ differs significantly from those using a fixed $U$. 
The increase of $U$ with strain magnitude can be understood as suppressing effective hopping (See Eq.~\ref{udmc}), thereby driving the strained system toward an orbital-overlap regime more similar to that of pristine monolayer MBT, where FM is favored over AFM.
This highlights the robustness of the FM state in monolayer MBT across a wide range of strain conditions. 
Combined with the weak strain dependence of the local moment, this robustness is consistent with the view that FM order may remain stable near the surface, where intralayer magnetic coupling dominates over interlayer coupling.
This, in turn, provides insight into the possible persistence of the quantum anomalous Hall effect (QAHE) in few odd-layer MBT structures under strain, where uncompensated magnetic moments are present.
While the required strain magnitudes exceed typical limits in bulk crystals, thin films can often sustain strains of 5-10$\%$ without structural degradation according to previous reports~\cite{gudelli2023strain,peng2020strain, xue2020control}.
Therefore, the robustness of the FM phase across extended strain ranges should be understood as particularly relevant in the thin-film context, highlighting its intrinsic stability under significant perturbations.

\section{Conclusion}
In this work, we used DFT+$U$ and DMC to examine how strain and correlation are intertwined in monolayer MBT. 
The main conclusion is that the magnetic response of this system cannot be described reliably by adopting a single empirical Hubbard $U$ over the full strain range. 
Instead, the DMC benchmark indicates that the effective optimal $U$ evolves systematically with strain and can be represented, to good approximation, by a simple isotropic quadratic form. 
Using this DMC-informed $U$ within PBE+$U$ substantially improves the description of Mn local moments relative to fixed-$U$ choices and also changes the resulting strain-dependent magnetic landscape.
Importantly, although the strain-dependent $U$ model is isotropic, its impact on the magnetic properties is not, because it modifies the strain dependence of the dominant exchange interactions. 
The resulting robustness of the FM phase in monolayer MBT across a wide strain range further provides insight into the possible persistence of the QAHE in few odd-layer MBT structures under strain, where uncompensated magnetic moments are present. 
More broadly, this work shows that monolayer MBT provides a useful setting for benchmarking correlation-sensitive magnetism and that many-body-guided calibration of DFT+$U$ offers a practical route toward more reliable predictions in magnetic van der Waals materials.

\clearpage

\section*{Acknowledgement}
This work was supported by the U.S. Department of Energy, Office of Science, Office of Basic Energy Sciences, Materials Sciences and Engineering Division (F.A.R., M.E., M.Y., S.G.) and by the U.S. Department of Energy (DOE), Office of Science, National Quantum Information Science Research Centers, Quantum Science Center (S.-H.K.). 
J.A. (DMC calculations, writing) and J.T.K. (mentorship, analysis, writing) were supported by the U.S. Department of Energy,
Office of Science, Basic Energy Sciences, Materials Sciences and Engineering Division, as part of the Computational Materials Sciences Program and Center for Predictive Simulation of Functional Materials.
This work (calculating magnetic anisotropy) was also partly supported by the Korean government (MSIT) through the National Research Foundation of Korea (NRF) (2022R1A2C1005505 and 2022M3F3A2A01073562) and Institute for Information \& Communications Technology Planning \& Evaluation (IITP) (2021-0-01580) (D.J., Y.-K.K.).  An award of computer time was provided by the Innovative and Novel Computational
Impact on Theory and Experiment (INCITE) program. This research used resources of the Oak Ridge Leadership Computing Facility at the Oak Ridge National Laboratory, which is supported by the Office of Science of the U.S. Department of Energy under Contract No. DE-AC05-00OR22725.  This research also used resources of the National Energy Research Scientific Computing Center, a DOE Office of Science User Facility supported by the Office of Science of the U.S. Department of Energy under Contract No. DE-AC02-05CH11231 (M.E. and S.G) and using NERSC award BES-ERCAP0024568. 

\bibliographystyle{apsrev}
\bibliography{biblio.bib}

\end{document}